\documentstyle[12pt,epsfig]{article}
\begin{document}
\begin{center}
\vspace{.2in}
{\LARGE
NEW EVIDENCE FOR TRITON AND

\vskip .5 cm

HELION CLUSTERING IN NUCLEI}
\end{center}
\vspace{.4in}
\begin{center}
{\bf Afsar Abbas}\\
\vspace{.05in}
Institute of Physics\\ 
Bhubaneshwar-751005, India\\
\vspace{,05in}
email: afsar@iopb.res.in
\end{center}
\vspace{.5in}
\begin{center}
{\bf Abstract}
\end{center}
\vspace{.3in}

Normally one plots separation energies
$S_{1n}$ and $S_{2n}$ as a function of neutron number $N$ for a
particular proton number $Z$ or plot $S_{1p}$ and $S_{2p}$ as a
function of $Z$ for a particular $N$.  
Here we plot the separation energies a little differently:
$S_{1n}$ and $S_{2n}$ as a function of $Z$ for a particular $N$.
The same with  $S_{1p}$ and $S_{2p}$ as a function of N for a particular
Z. A systematic study of these new plots brings out certain very 
interesting generic features which clearly indicate presence 
of triton ("t")  ${^{3}_{1} H_{2}}$
and helion ("h")  ${^{3}_{2} He_{1}}$ clustering in nuclei.

\newpage

In the study of nuclear structure, separation energies have played 
an important role and are continuing to do so in identifying
new magic numbers in recent years [1-4]. In all such studies one
normally plots separation energies
$S_{1n}$ and $S_{2n}$ as a function of neutron number $N$ for a
particular proton number $Z$ or plot $S_{1p}$ and $S_{2p}$ as a
function of $Z$ for a particular $N$.

We plot the separation energies a little differently here.  
We plot
$S_{1n}$ and $S_{2n}$ as a function of $Z$ for a particular $N$.
The same with  $S_{1p}$ and $S_{2p}$ as a function of N for a particular
Z. We do a sytematic study of these plots for all the data 
available in literature at present [5]. 
Such plots have completely been ignored in literature.
This is very unfortunate as these plots do contain huge amount
of interesting and revealing information which has been missed so 
far. We shall show that these brings out certain very 
interesting generic features which as we shall find clearly 
indicate strong evidences of   
triton ("t")  ${^{3}_{1} H_{2}}$
and helion ("h")  ${^{3}_{2} He_{1}}$ 
clustering in nuclei

We study the whole range of data set available [5]. 
However as all this
will take much more space than necessary for this paper, here we 
show a few representative plots. These are 
one and two neutron separation energy as a function of Z 
for N = 29 and 30 as plotted in Fig 1 and 2 below.
All the plots of 
$S_{1n}$ and $S_{2n}$ as a function of $Z$ for a particular $N$
show similar features.
The same with  $S_{1p}$ and $S_{2p}$ as a function of N for a particular
Z. We do a sytematic study of these plots for all the data.
The features which we shall point out here are there in all the 
other plots. In fact we shall study here only those features which 
are generic of all such plots. Special studies of specific
set of nuclei ( which too would be very fruitful ) shall be 
left for separate future studies.

Certain common feautres which stand out are as follows
(The statements below are made in the context of
the plot
$S_{1n}$ and $S_{2n}$ as a function of $Z$ for a particular $N$):

\vskip .5 cm

{\bf A}. {\it For all even-even N=Z nuclei there is always a 
pronounced larger separation energy required with respect to the 
lower adjoining nuclei plotted}.

\vskip .5 cm

{\bf B}. {\it For all odd-odd N=Z nuclei there is always a 
pronounced larger separation energy required with respect to the 
lower adjoining nuclei plotted}.

\vskip .5 cm

{\bf C}. {\it For case {\bf A} when Z number is changed by one 
unit, the separation energy hardly changes (sometimes not at all). 
But when this number is changed by two units, another 
pronounced peak occurs}.

\vskip .5 cm 

{\bf D}. {\it For case {\bf B} when Z  number 
is changed by one unit, the 
separation energy hardly changes (sometimes not at all). But when 
this number is changed by two units, another pronounced peak occurs.
So there are peaks for odd-odd N-Z nuclei ( and not for 
even-even cases )}.

\vskip .5 cm 

Note that for example here we are plotting 
$S_{1n}$ and $S_{2n}$ as a function of $Z$ for a particular $N$.
Hence while we are pulling one or two N we are studying this
as fuction of Z. Thus the above effects cannot be the result of
identical nucleon pairings, What these plots are telling us is as
to what happens to last one or two neutron bindings in a nucleus 
as proton number changes.

Clearly the peaks as indicated in {\bf A} 
above are due to the fact 
that the last one or two neutrons must have come from a stable 
alpha 
cluster. This is consolidated by the fact that another extra Z 
does not make a difference to the saparation energy.

To understand observation {\bf B} 
above - note that here the
peaks are there for ALL odd-odd N=Z nuclei. This is an amazing
fact. These odd-odd nuclei are more stable or "magic" with respect 
to the adjoining odd-even or even-odd nuclei.
Pairing of identical nucleons cannot explain this generic 
feature. Neither can alpha clustering do so.
Obviously it is the formation of triton-helion 'h-t' pair
which can only explain this extraordinary effect.
This is the minimal requiremet without which one just cannot 
explain the generic effect indicated in 'B'.
Just as alpha clustering explains the extra stability for case 
{\bf A}
so does the formation of 'h-t' cluster pair
can only exlain the data in {\bf B}. 

What I would like to 
emphasize here is that the above conclusion is inescapable as per 
the empirical information manisfested in separtation energy 
plotted in the manner indicated above.
It is Nature which is forcing this conclusion upon us.

To understand observations {\bf C} and {\bf D} we have to 
undertand new empirical studies on neurton rich nuclei.

Going through the binding energy systematics of neutron rich nuclei one
notices that as the number of 
$ \alpha $'s increases along with the neutrons, each $ ^{4}He $ + 2n pair
tends to behave like a cluster of two 
$ ^{3}_{1}H_{2} $ nuclein [6]. 
Hence the author had concluded that
all light neutron rich nuclei $ _{Z}^{3Z}A_{2Z} $ are made up of Z 
number of $ ^{3}_{1}H_{2} $ clusters. 
There are good empirical evidence for this.

Separation energy studies like above support this conclusion 
too [7].
We plotted separation energies for neutron number N and for proton 
number Z
fixed separately at 4, 6, 8, 10, 11, 12, 16, 20, 22 and 24.
We found extra-ordinary stabilty manifested 
by the plotted data for the proton and neutron pairs (Z,N): 
(6,12), (8,16), (10,20), (11,22) and (12,24). 
These new magicities were present in the neutron rich sector for 
the pair (Z,N) where N=2Z.

What is the significance of this extraordianry stabilty or magicity for
all the nuclei ${^{3Z}_Z} A _{2Z}$? 
Quite clearly the only way we can explain the extra magicity for these
N=2Z nuclei is by invoking the significance of triton clustering
in the ground state of these neutron rich nuclei. 
So the nucleus ${^{3Z}_Z} A _{2Z}$ is made up of Z number of 
triton clusters.

To understand this unique feature the author introduced a new
symmetry "nusospin" symmetry [7]. Though the thrust of this paper
is not this nusospin summetry, the author would like to emphasize 
that triton and helion clustering effects in nuclei can be 
understood naturally in the framework of the new nusospin 
symmetry.

So it is clear that it is tritons which explain the stabilty of 
neutron rich nuclei and it it is pair of 'h-t' clusters which 
explain the stabilty of odd-odd N=Z nuclei in the separation 
energy as plotted above.

Now we can explain the obseravtion {\bf C} above. Clearly for 
even-even N=Z nuclei it is one ( or more ) alpha clusters which 
explain the data. Hence one extra Z does not affect the 
separation energy. But two extra Z will tend to make a
pair of helions ( akin to the two tritons for neutron rich 
case above ).
These two 'h-h' will make for the extra stabilty for the adjoing 
even-even nuclei (and so on). 
So also can the observation {\bf D} be 
understood as the extra 2Z will create an extra helion to attach 
to the already existing 'h-t' pair to make for extra stabilty 
of this adjoining odd-odd nuclei. Other peaks in the above plots 
can be similarly explained as due to alpha or triton and helion 
clusters. 

The author would like to emphasize that
quite obviusly these qualitative conclusions are 
inescapable and unique. 
This is true as there is no other way to explain the above 
empirical observations on the whole.

\newpage

\begin{figure}
\caption{One and two neutron separation energy as a function of 
proton number Z for fixed N=29 neutrons}
\epsfclipon
\epsfxsize=0.99\textwidth
\epsfbox{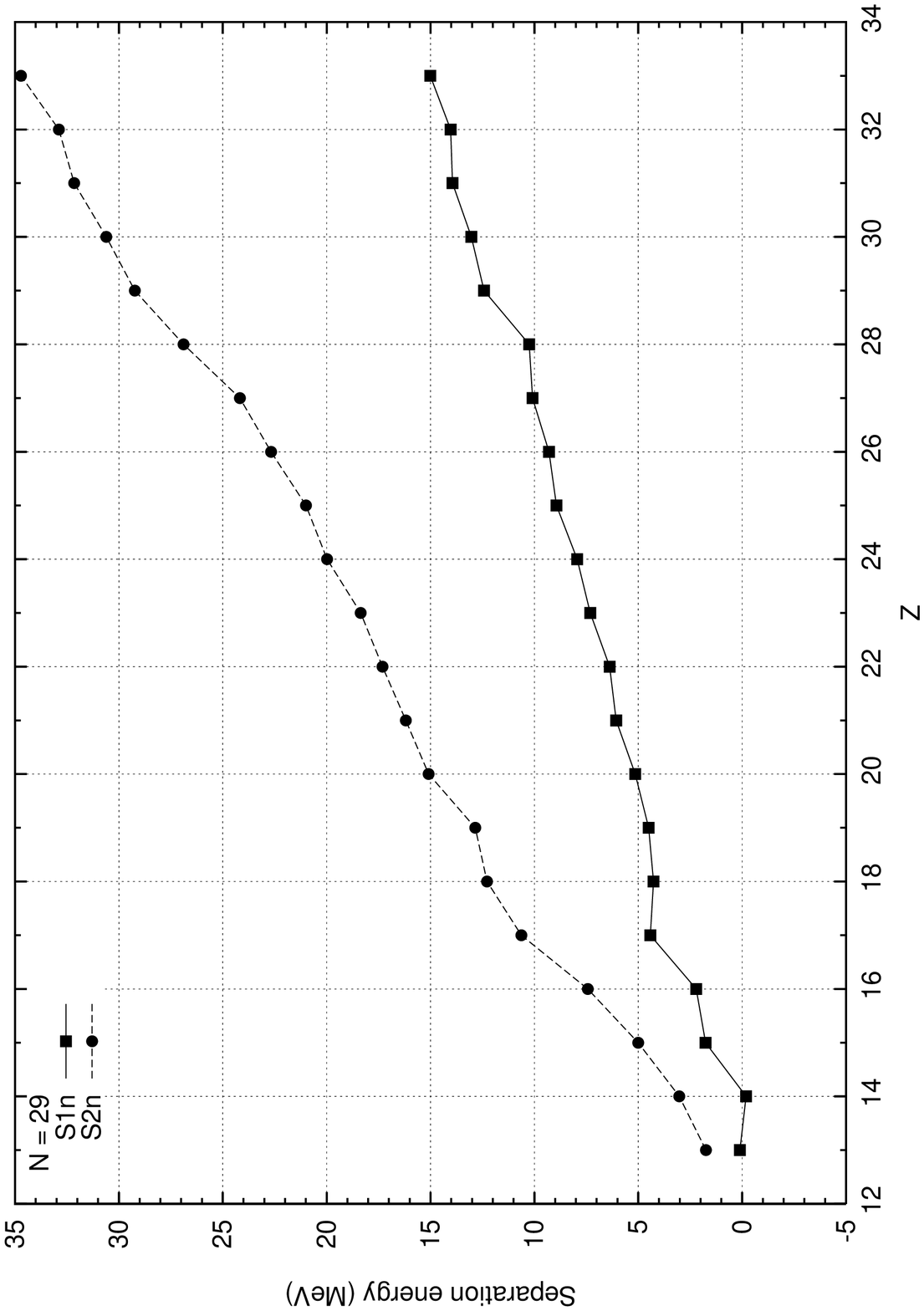}
\end{figure}

\newpage

\begin{figure}
\caption{One and two neutron separation energy as a function of 
proton number Z 
for fixed N=30 neutrons}
\epsfclipon
\epsfxsize=0.99\textwidth
\epsfbox{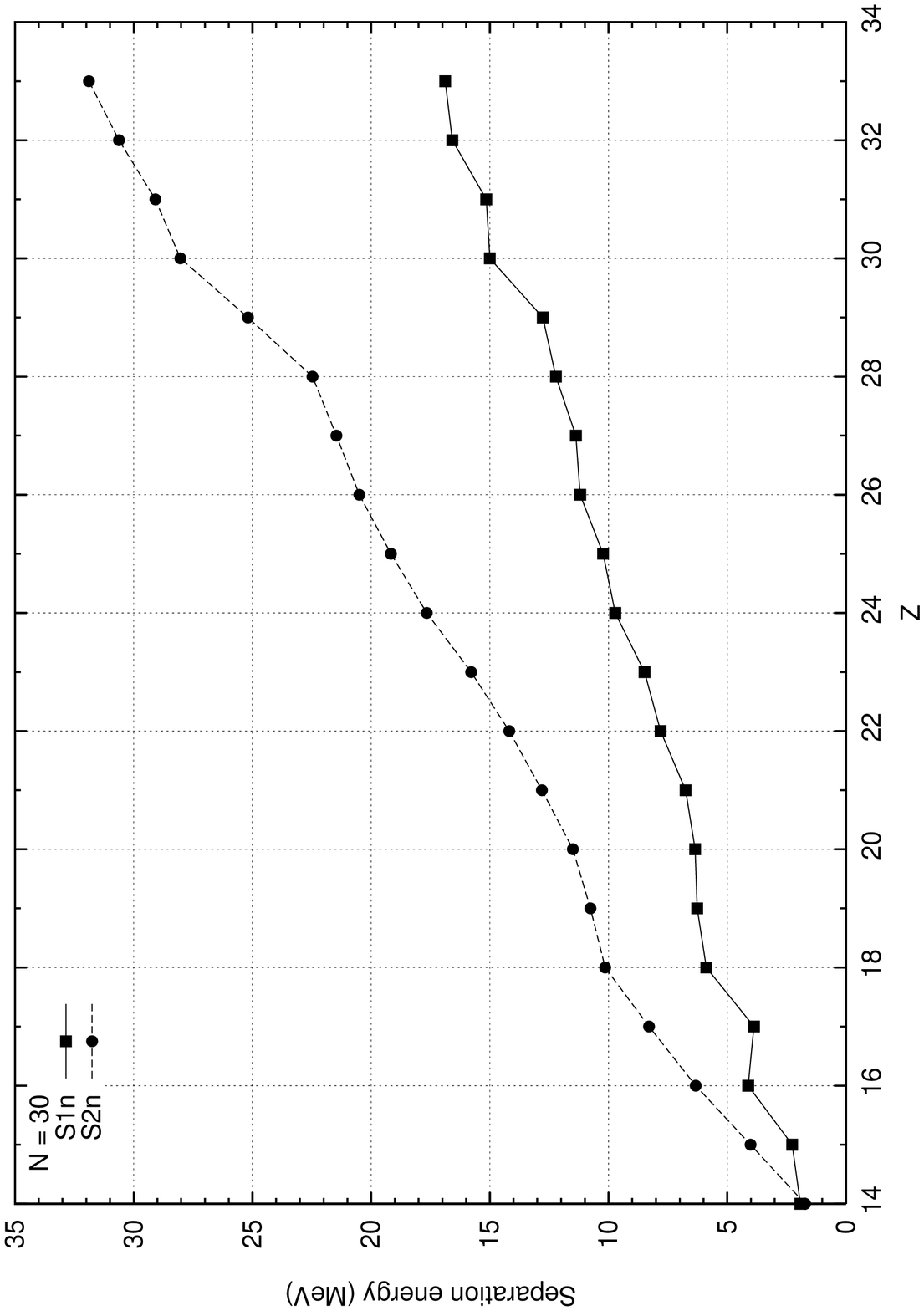}
\end{figure}

\newpage

\vskip .3 cm

\begin{center}
{\bf REFERENCES }
\end{center}
\vspace{.4in}

1. I. Tanihata, Nucl. Phys. {\bf A682}, (2001) 114c

2. R. Kanungo, I. Tanihata and A. Ozawa, 
Phys. Lett.  {\bf B 528} (2002) 58

3. Z. Dlouhy, D. Baiborodin, J. Mrazek and G.
Thiamova, Nucl. Phys. {\bf A722} (2003) 36c

4. M. Thoennessen, T. Baumann, J. Enders, N. H.
Frank, P. Heckman, J. P. Seitz and E. Tryggestad, Nucl. Phys.
{\bf 26} (2003) 61c

5. G. Audi, A. H. Wapstra and C. Thibault, Nucl. Phys.
{\bf A 729} (2003) 337

6. A. Abbas, Mod. Phys. Lett. {\bf A 16} (2001) 755

7.  A, Abbas, Mod. Phys. Lett. {\bf A 20} (2005) 2553

\end{document}